\ifx\mnmacrosloaded\undefined \input mn\fi
%

\newif\ifAMStwofonts
\ifCUPmtplainloaded \else
  \NewTextAlphabet{textbfit} {cmbxti10} {}
  \NewTextAlphabet{textbfss} {cmssbx10} {}
  \NewMathAlphabet{mathbfit} {cmbxti10} {} 
  \NewMathAlphabet{mathbfss} {cmssbx10} {} 
  \ifAMStwofonts
    \NewSymbolFont{upmath} {eurm10}
    \NewSymbolFont{AMSa} {msam10}
    \NewMathSymbol{\upi}     {0}{upmath}{19}
    \NewMathSymbol{\umu}     {0}{upmath}{16}
    \NewMathSymbol{\upartial}{0}{upmath}{40}
    \NewMathSymbol{\leqslant}{3}{AMSa}{36}
    \NewMathSymbol{\geqslant}{3}{AMSa}{3E}

     \let\ge=\geqslant
  \else
    \def\umu{\mu}
    \def\upi{\pi}
    \def\upartial{\partial}
  \fi
\fi


\pageoffset{-1.5pc}{2pc}
\input psfig.tex

\loadboldmathnames



\pagerange{000--000}    
\pubyear{1996}
\volume{000}

\def\la{\mathrel{\hbox{\rlap{\hbox{\lower4pt\hbox{$\sim$}}}\hbox{$<$}}}}
\def\ga{\mathrel{\hbox{\rlap{\hbox{\lower4pt\hbox{$\sim$}}}\hbox{$>$}}}}

\def\ax{$\alpha_{\rm x}$}
\def\ea{et al.\ }

\begintopmatter  

\title{The ROSAT X-ray spectra of BL Lacertae objects} 
\author{Paolo Padovani$^1$ and Paolo Giommi$^2$}

\affiliation{$^1$ Dipartimento di Fisica, II Universit\`a di Roma ``Tor
Vergata'', Via della Ricerca Scientifica 1, I-00133 Roma, Italy}
\medskip
\affiliation{$^2$ SAX, Scientific Data Center, ASI, Viale Regina Margherita 202,
I-00198 Roma, Italy} 

\shortauthor{P. Padovani and P. Giommi}

\shorttitle{X-ray spectra of BL Lacs}


\acceptedline{Accepted 1995 October 11. Received 1995 July 31}

\abstract {We study the X-ray spectra of 85 BL Lacertae objects using the
hardness ratios as given in the WGA catalogue of {\it ROSAT} sources. This
sample includes all WGA BL Lacs with high-quality data and comprises about 50
per cent of presently known BL Lacs. We find that BL Lacs have energy
power-law spectral indices between 0 and 3 with a mean value \ax~$\sim 1.4$.
Significant differences, however, are present between high-energy cutoff BL
Lacs (HBLs), normally selected in the X-ray band, and low-energy cutoff BL
Lacs (LBLs), generally found in radio surveys. HBLs have steeper X-ray
spectral slopes (\ax~$\sim 1.5$) well correlated with $\alpha_{\rm ox}$, and
anti-correlated with the X-ray-to-radio flux ratio and cutoff frequency, with
convex overall broad-band spectra. LBLs, on the other hand, have flatter X-ray
spectra (\ax~$\sim 1.1$) and concave optical-X-ray continuum. We interpret
these results in terms of different mechanisms being responsible for the X-ray
emission in the two classes, namely synchrotron and inverse Compton for HBLs
and LBLs respectively. The observed differences are consistent with the
hypothesis that HBLs and LBLs are powered by essentially the same non-thermal
engines differing mainly in their synchrotron cutoff energy.} 


\maketitle  

\section{Introduction} 

BL Lacertae objects constitute a relatively rare class of active galactic
nuclei (AGN), distinguished by their almost complete lack of emission lines.
At present, only about two hundred BL Lacs are known (Padovani \& Giommi 1995b)
as compared, for example, to the thousands of quasars listed in the most
recent AGN catalogues (V\'eron-Cetty \& V\'eron 1993; Hewitt \& Burbidge
1993). 

Despite their small numbers and other somewhat extreme and non-typical
properties (radio core-dominance, superluminal motion, rapid variability, high
polarization: e.g., Kollgaard 1994; Urry \& Padovani 1995) BL Lacs are
nevertheless an important AGN subgroup, as the broad-band emission in these
objects is most likely completely dominated by non-thermal processes. Their
multifrequency spectra are in fact fairly smooth over a wide frequency range,
suggestive of a common emission process, and lack the thermal features
characteristic of the other, more common, AGN, like dust emission in the
infrared and (possibly) disk emission in the ultraviolet band (e.g., Bregman
1990). 

Historically, BL Lacertae objects have been divided into RBLs (radio-selected
BL Lacs) and XBLs (X-ray-selected BL Lacs), depending on the discovery band. 
The two classes showed in fact somewhat different properties, RBLs being 
more extreme, for example, in their optical polarization, core-dominance, and
variability properties (e.g., Jannuzi, Smith \& Elston 1994; Perlman \& Stocke
1993). This division was clearly not satisfactory, as it was not based on
intrinsic physical properties but solely on the selection band, and did not
characterize uniquely a source. Recent all-sky X-ray surveys
(Laurent-Muehleisen \ea 1993; Brinkmann \ea 1995; Perlman \ea 1995a), in fact,
include many BL Lacs previously selected at radio frequencies, which could now
then be classified both as XBLs or RBLs. 

It has been known for some time (e.g., Ledden \& O'Dell 1985) that the
broad-band spectra of the two classes are also different (with the likely
exception of the radio to millimetre band: Gear 1993). Recently, Giommi,
Ansari \& Micol (1995a) have studied the multifrequency spectrum of a large
sample of BL Lacs, arguing that there may be only one population of objects,
characterized by a wide range of energy cutoffs (or alternatively energy peaks
in a $\nu f_{\nu}$ vs. $\nu$ plot): at infrared-optical energies for most
RBLs, at ultraviolet/X-ray or higher energies for most XBLs, with the latter 
constituting only $\sim 10$ per cent of all BL Lacs. The X-ray-to-radio flux
ratios ($f_{\rm x}/f_{\rm r}$) for the two classes are also different, as a
result of the different break frequencies. Note that within this hypothesis
there is a continuum of cutoff energies, with different selection bands
picking up objects with different spectra. 

We (Giommi \& Padovani 1994; Padovani \& Giommi 1995a) have then proposed a
division based on $f_{\rm x}/f_{\rm r}$ (as the break frequency at present can
be derived directly from the spectra only for a small number of objects: see
Sect. 3), classifying as HBLs (or high-energy cutoff BL Lacs) and LBLs (or
low-energy cutoff BL Lacs) sources respectively above or below the value of
$f_{\rm x}/f_{\rm r} \sim 10^{-11}$ (where X-ray fluxes cover the 0.3 -- 3.5
keV range and are in units of erg cm$^{-2}$ s$^{-1}$ and radio fluxes refer to
5 GHz and are expressed in janskys). We have also shown that the hypothesis
that HBLs (that is, so far, most XBLs) are the minor component of the BL Lac
population with high-energy cutoffs can explain most of their properties at
least as well as the alternative (and more diffuse) explanation for the
existence of two classes of BL Lacs, which posits that XBLs are intrinsically
more numerous as they are viewed significantly off their beaming axis (e.g.,
Stocke \ea 1985; Maraschi \ea 1986; Padovani \& Urry 1990). 

A study of the puzzling distribution of BL Lacs on the $\alpha_{\rm ro},
\alpha_{\rm ox}$ plane (Padovani \& Giommi 1995a) has led us to suggest that
the observed bimodality can also be explained by an energy break ranging from
high (X-ray) to low (far-infrared) energies for HBLs and LBLs respectively. 
This assumes a
different origin of the X-ray emission for HBLs and LBLs: namely, an extension
of the synchrotron emission likely responsible for the lower energy continuum
in the former and inverse Compton (SSC: synchrotron self-Compton) emission in
the latter. This would be consistent with the peak of the emitted power being
at lower energies for LBLs, suggestive of an extra component taking over at
higher frequencies (see Bregman 1990 and references therein). A crucial test
of our hypothesis (and in general of any other model wishing to explain the
existence of two BL Lac classes) is then the analysis of the spectral shape of
the X-ray emission of a large sample of BL Lacs including both HBLs and LBLs.
In the X-ray band, in fact, Compton emission is expected to be flatter than
the synchrotron component (e.g., Ghisellini \& Maraschi 1989). 

The study of X-ray spectral indices of BL Lacs has given various and sometimes
contradicting results, even considering only the most recent papers dealing
with possible differences between the two classes (references to earlier
papers can be found in Sambruna \ea 1994). Sambruna \ea (1994) have analyzed
the EXOSAT spectra of 21 BL Lacs (only five of which were RBLs) and found
similar spectral indices in the LE ($0.04-2.0$ keV) and ME ($1-20$ keV) bands
for the two classes, with indications of a steepening at higher energies for
XBLs. Ciliegi, Bassani \& Caroli (1995) have studied the X-ray spectra of 42
BL Lacs (mostly RBLs) observed with various (pre-{\it ROSAT}) X-ray telescopes
in different energy ranges and found that XBLs and RBLs have similar (within
the rather large errors) spectral indices in the soft ($0.2-4$ keV) and
possibly also in the hard ($2-10$ keV) band. However, there were indications
that XBLs and RBLs have different global X-ray spectra, with the former and
the latter steepening and flattening, respectively, at higher energies. Urry
\ea (1996; see also Sambruna 1994) have examined more recent {\it ROSAT}
spectral data for 32 out of the 34 radio-selected BL Lacs in the 1 Jy sample
(Stickel \ea 1991). They find that 1 Jy RBLs have spectra similar to those of
the {\it Einstein Observatory} Extended Medium Sensitivity Survey (EMSS) XBLs
(Perlman \ea 1995b), with an indication for the subsample with {\it Einstein}
IPC data of a flattening of the spectra at higher ($\ga 1$ keV) energies.
Comastri, Molendi \& Ghisellini (1995) analyzed the {\it ROSAT}
spectra of twelve 1 Jy BL Lacs (i.e. a subsample of the objects studied by
Urry \ea 1996) dividing the objects according to the X-ray-to-radio flux
ratio, as proposed by Padovani \& Giommi (1995a), with the result that the
average spectrum of the (3) HBLs was steeper than that of the (9) LBLs. Urry 
\ea (1996) reached similar conclusions. 

It is clear that at least some of the different results can be explained by 
poor statistics and the use of the RBL/XBL classification, which as we have 
discussed above is not physical. However, it is also probably true that the
differences between the X-ray spectra of HBLs and LBLs are not that strong and
require better statistics to be singled out. 

The purpose of this paper is to analyze the X-ray spectra of all BL Lacs
observed (as pointed or serendipitous sources) by {\it ROSAT} to check for any
differences between HBLs and LBLs, coupling for the first time a large
statistics with what we think is a meaningful distinction between the two
classes. Our data have been taken from the WGA catalogue (White, Giommi \&
Angelini 1994), a large list of X-ray sources generated from all the {\it
ROSAT} PSPC pointed observations, with which is now possible to study the
X-ray properties of large numbers of objects in an homogeneous and relatively
simple way. The selection of the objects was done by cross-correlating the
first revision of the WGA catalogue (restricting ourselves to sources with
quality flag $\ge 5$, which excludes problematic detections) with our recent
catalogue of BL Lacs (Padovani \& Giommi 1995b). This produced 163
observations of 85 distinct BL Lacs (58 HBLs and 27 LBLs: see Sect. 2), which
correspond to $\simeq 46$ per cent of confirmed BL Lacs presently known. This
represents the largest number of BL Lacs for which homogeneous X-ray spectral
information is available and the largest BL Lac sample ever studied at X-ray
frequencies (cf. Ciliegi et al. 1993 who collected information on the spectra
of 42 BL Lacs observed with various X-ray telescopes). 

The structure of the paper is as follows: Section 2 describes the
observational data and the data analysis, Section 3 studies the X-ray spectral
properties of BL Lacs while Section 4 discusses our results and presents our
conclusions. Throughout this paper spectral indices are written $f_{\nu}
\propto \nu^{-\alpha}$. 

\section{Data Analysis}

Spectral indices for the WGA BL Lacs were obtained from the count rates given 
in the catalogue in the $0.1-0.4$ keV range (soft band: $S$), $0.4-0.86$ keV
range (mid band: $M$) and $0.87-2.0$ keV range (hard band: $H$) combined into
hardness ($HR$) and softness ($SR$) ratios (see Giommi et al. 1995b). Briefly,
the count rates were combined to construct one $SR = S/M$ and two HRs, $HR_1 =
H/M$ and $HR_2 = H/(M+S)$. Initially we converted the hardness ratios into
energy spectral indices both assuming Galactic $N_{\rm H}$ derived from 21 cm 
measurements (Stark et al. 1992; Shafer et al., private communication) 
and with $N_{\rm H}$ derived from
the softness ratio through an iterative procedure. The spectral indices
derived from the two approaches were found to be similar and well correlated
in the $0.4-2.0$ keV range, while considerable scatter was present when the
total {\it ROSAT} band was considered, with a sizeable number of objects
having $\alpha_{\rm x,NH}$ significantly different from $\alpha_{\rm
x,Galactic~NH}$, suggestive of a change of the spectrum at lower energies. In
this paper we will adopt as X-ray spectral index the value obtained with
$N_{\rm H}$ fixed to the Galactic value, which is better determined than the
$N_{\rm H}$ derived from the softness ratios, and in the $0.4-2.0$ keV range,
i.e. the mid to hard band (spectral indices in the whole {\it ROSAT} range
were used for the comparison between the values obtained with our method and
those derived by other authors from detailed spectral fits: see below). 

The choice of this energy range stems from two facts: first, our method {\it
assumes} a single power-law; given the evidence of a change of spectral shape
at lower energies in some sources, as described above (see also Comastri \ea 
1995 and Urry \ea 1996), this assumption might be incorrect in the whole {\it
ROSAT} range while it is probably more acceptable in the narrower $0.4-2.0$ keV
range (as shown by the similarity of the spectral indices obtained with
Galactic $N_{\rm H}$ and $N_{\rm H}$ derived from the softness ratio); second,
the method used in the WGA catalogue to estimate the source intensity uses the
counts detected in a box whose size optimises the signal to noise (S/N) ratio.
This size is calculated assuming an average point spread function (PSF) that
is too sharp for very soft photons. For weak sources near the field center the
``optimum'' box size is small and consequently includes a small,
energy-dependent, fraction of the source photons causing an underestimation of
the counts in the soft band. 
The effects of the dependence of energy response with off-axis angle have been
taken into account using different conversion matrices in five different
off-axis ranges: $0-20$, $20-30$, $30-40$, $40-50$, and $50-60$ arcminutes. 

Errors on the spectral indices ($1 \sigma$) were derived from the
uncertainties on the hardness ratios. We included in our analysis only
observations with values of hardness ratios relative to their uncertainties $>
3.5$, which corresponds roughly to $1\sigma$ errors on $\alpha_{\rm x} \la
0.5$ (Giommi et al. 1995b). All objects in this study are ``bona fide'' BL
Lacs: we excluded eight sources classified as uncertain BL Lacs or BL Lac
candidates (Padovani \& Giommi 1995b). We are then left with 163 observations
of 85 BL Lacs. 

\beginfigure{1}
\psfig{file=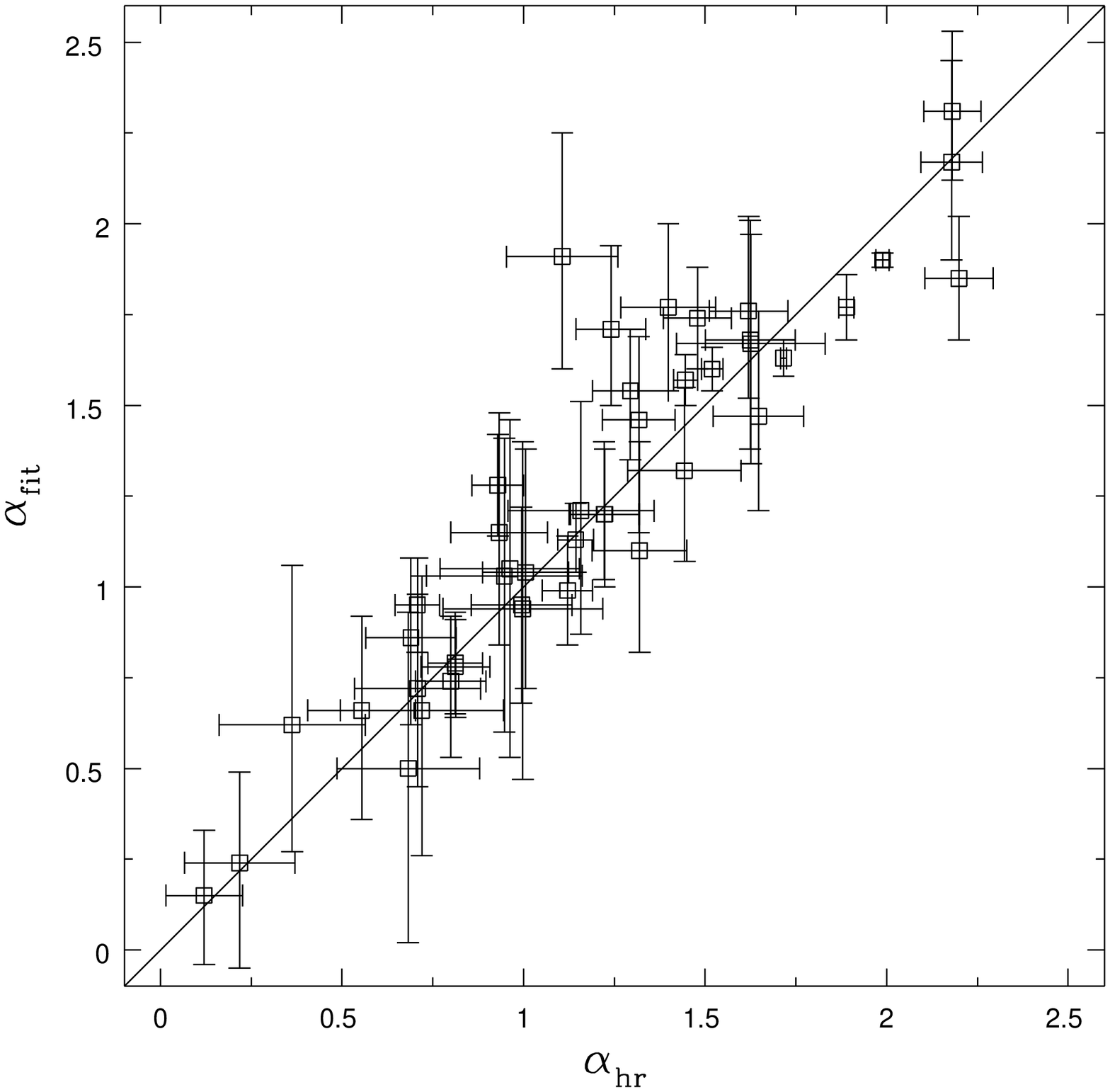,height=8.5truecm,width=8.5truecm}
\caption{{\bf Figure 1.} {\it ROSAT} band energy spectral indices of BL Lacs
derived from our approach based on the hardness ratios, $\alpha_{\rm hr}$,
compared to the values obtained by other authors from a detailed spectral
fitting procedure of the same {\it ROSAT} data, $\alpha_{\rm fit}$. The solid
line indicates the locus of points with $\alpha_{\rm hr} = \alpha_{\rm fit}$.
Error bars represent $1\sigma$ errors for our values and 90 per cent errors
for those obtained by other authors.} 
\endfigure

Figure 1 compares the X-ray energy spectral indices derived from our method in
the whole {\it ROSAT} band (i.e. obtained from $HR_2$) with those obtained by
other authors (Perlman et al. 1995b; Maraschi et al. 1995; Urry et al. 1996)
through a detailed spectral fitting procedure, using the same {\it ROSAT} data
for 44 BL Lacs. The comparison is between spectral indices derived under the
same assumption regarding the absorbing column density, i.e. either assuming
Galactic $N_{\rm H}$ (Stark et al. 1992; Shafer et al., private communication) 
or with $N_{\rm H}$
derived from the softness ratio. Figure 1 shows that most objects cluster
around the line $\alpha_{\rm x,hr} = \alpha_{\rm x,fit}$ with a linear
relationship between the spectral indices derived with the two methods.
Overall the agreement is then very good, especially considering the
underestimate of soft photons in the WGA data. The mean values for the two
derivations are $\alpha_{\rm x,hr} = 1.18\pm0.08$ and $\alpha_{\rm x,fit} =
1.24\pm0.08$ (here and in the following we give the error of the mean), which
clearly indicates the statistical robustness of our approach. 
 
A few sources in the PSPC field of famous objects have many X-ray
observations: for example, 1E 1415+2557 (in the field of NGC 5548) has been
observed in 29 different occasions! Inclusion of all these data in statistical
studies (means, correlations, Kolmogorov-Smirnov [KS] tests etc.) would
clearly bias them: the resulting mean, for example, would represent more the
typical X-ray spectrum of 1E 1415+2557 than that of BL Lacs as a class.
Therefore, although we plot in the figures all the 163 X-ray spectral indices,
in all statistical tests we have kept only one observation (and therefore one
spectral index) per object, selected on a one-to-one basis taking the best
combination of offset from the field center and S/N ratio. When more than
two observations with comparable offset and S/N ratios were available, we
selected as most representative the one with \ax~nearest to the mean value
for that source. 


\section{The X-ray spectra}

\subsection{\ax~versus $f_{\rm x}/f_{\rm r}$}  

The main purpose of this paper is to study the X-ray spectral indices of HBLs
and LBLs. To this aim, it is important to define the value of the $f_{\rm
x}/f_{\rm r}$ ratio which divides the two classes. (We prefer the use of
broad-band X-ray fluxes in the X-ray-to-radio flux ratios instead of, for
example, 1 keV fluxes, because the latter are more dependent on the precise
value of \ax~and therefore require spectral fitting, which is at present not
possible for all BL Lacs.) In Padovani \& Giommi (1995a), based on the energy
distributions given by Giommi \ea (1995a), we suggested a ratio $f_{\rm
x}/f_{\rm r} \sim 10^{-11}$. It should be noted, though, that very few BL Lacs
in the compilation by Giommi et al. (1995a) had $10^{-11.5} \la f_{\rm 
x}/f_{\rm r} \la 10^{-10.5}$, because this intermediate zone is not favoured
either by radio or X-ray selection and is therefore poorly populated both in
X-ray and radio surveys. The dividing $f_{\rm x}/f_{\rm r}$ value was then not
well determined. 

\beginfigure{2}
\psfig{file=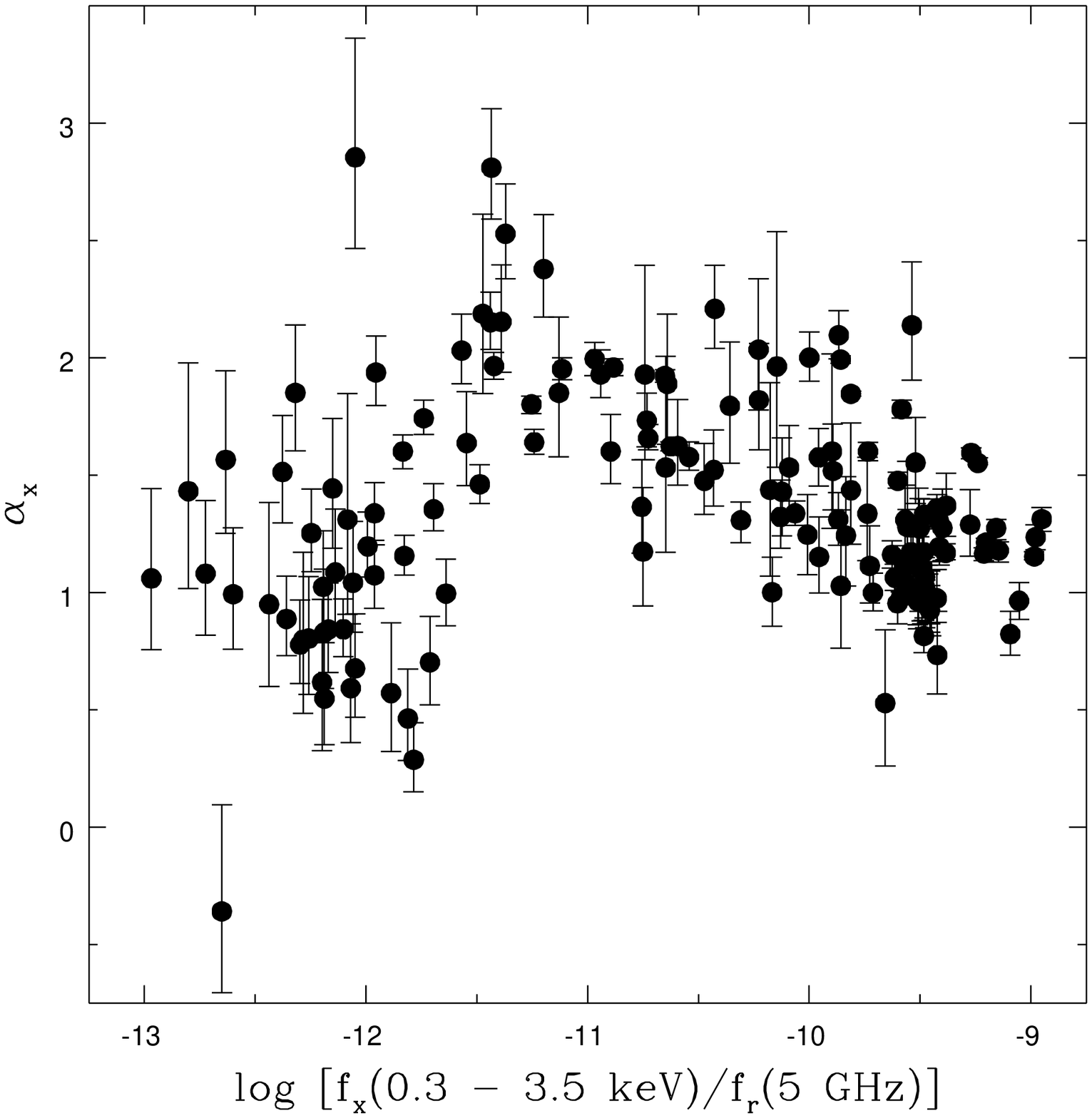,height=8.5truecm,width=8.5truecm}
\caption{{\bf Figure 2.} The X-ray spectral index of BL Lacs versus their
X-ray-to-radio flux ratio (with X-ray fluxes in the 0.3 -- 3.5 keV range in
units of erg cm$^{-2}$ s$^{-1}$ and radio fluxes at 5 GHz expressed in
janskys). Error bars represent $1\sigma$ errors.} 
\endfigure

In Fig. 2 we plot $\alpha_{\rm x}$ versus $f_{\rm x}/f_{\rm r}$ for all the
objects in our sample, with radio data from Padovani \& Giommi (1995b and
references therein) and X-ray fluxes in the 0.3 -- 3.5 keV band derived
from the Galactic $N_{\rm H}$ and \ax~values. (Ideally the radio and X-ray
observations should be simultaneous; although variability in the radio is not
very strong, this undoubtedly introduces some ``noise'' in our plots.) Within
our hypothesis on the difference between HBLs and LBLs (Padovani \& Giommi
1995a), described in the Introduction, the X-ray emission of HBLs should be
dominated by synchrotron emission. HBLs near the transition region, in which
the break is just below the X-ray band and where we see the steep tail of the
synchrotron emission but not yet the inverse Compton component, should then be
characterized by the steepest X-ray spectra (see also Sect. 3.3). Figure 2
shows this to happen for $10^{-11.5} \la f_{\rm x}/f_{\rm r} \la 10^{-11}$,
where all objects have $\alpha_{\rm x} \ga 1.5$ and reach values up to $\sim
3$. A better dividing line between the two classes is then $f_{\rm x}/f_{\rm
r} \sim 10^{-11.5}$, which is the value at which there is a {\it
discontinuity} in \ax. We stress that our results (and those obtained in 
Padovani \& Giommi 1995a) are unaffected by the
precise value of this parameter. (One object has \ax~$\sim 3$ but $f_{\rm
x}/f_{\rm r} \sim 10^{-12}$. This is 1WGA J1202.1+444, which appears three
times in the WGA catalogue and only in its faintest state crosses the HBL/LBL
boundary to become an LBL with an unusually steep X-ray spectrum, albeit with
large errors, following the \ax~$- f_{\rm x}$ anti-correlation normally
observed in HBLs [e.g., Giommi \ea 1990; Sambruna \ea 1994]. We note however
that a decrease in radio flux by a factor of $2-3$ as compared to the value we
used, which we cannot rule out, would have kept this object within the HBL
limits.) 

The mean value of the X-ray spectral index for the 85 BL Lacs under study is
$\langle \alpha_{\rm x} \rangle = 1.37\pm0.05$. Figure 2, however, already
shows that the X-ray spectral indices of the 58 HBLs (BL Lacs with $f_{\rm
x}/f_{\rm r} \ge 10^{-11.5}$) are {\it steeper} than those of the 27 LBLs (BL
Lacs with $f_{\rm x}/f_{\rm r} < 10^{-11.5}$). In fact, $\langle \alpha_{\rm
x,HBL} \rangle = 1.52\pm 0.06$ while $\langle \alpha_{\rm x,LBL} \rangle =
1.06\pm 0.09$, different at the 99.99 per cent level according to a Student's
t-test. The two \ax~distributions, shown in Fig. 3, are also different at the
99.9 per cent level according to a KS test. (Had the dividing line been drawn
at the value $f_{\rm x}/f_{\rm r} = 10^{-11}$, against the evidence presented
in Fig. 2, one would have found $\langle \alpha_{\rm x,HBL} \rangle = 1.43\pm
0.05$ and $\langle \alpha_{\rm x,LBL} \rangle = 1.27\pm 0.11$, with the two
distributions different at the 95.4 per cent level according to a KS test.) 

\beginfigure{3}
\psfig{file=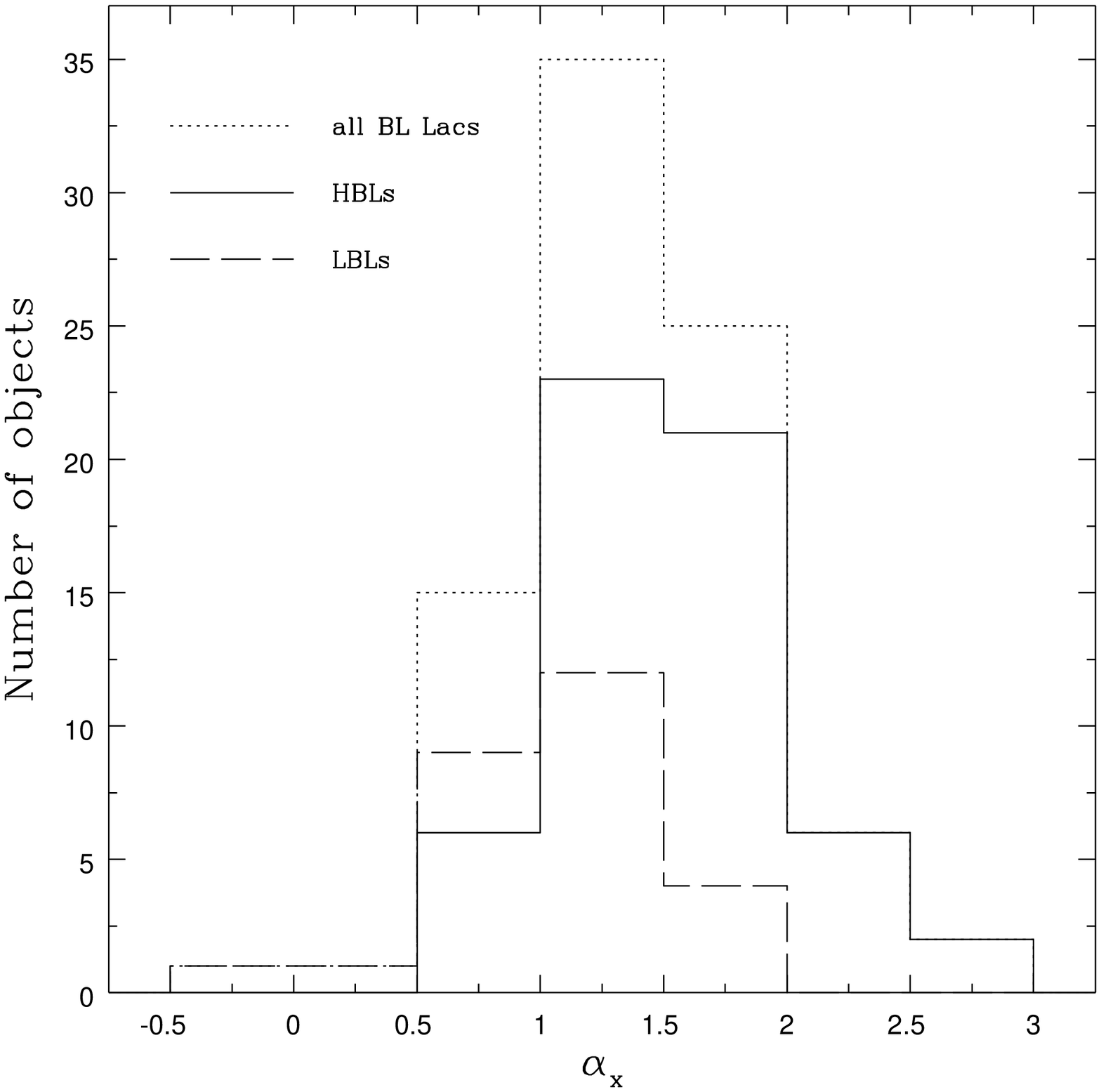,height=8.5truecm,width=8.5truecm}
\caption{{\bf Figure 3.} The X-ray spectral index distribution of all BL Lacs
in our sample (dotted line), and of HBLs ($f_{\rm x}/f_{\rm r} \ge
10^{-11.5}$: solid line) and LBLs ($f_{\rm x}/f_{\rm r} < 10^{-11.5}$:
dashed line) separately. Only one X-ray spectrum per object is used, as
described in Sect. 2. The distributions of HBLs and LBLs are different at the
99.9 per cent level according to a KS test.} 
\endfigure

Another notable feature of Fig. 2 is the anti-correlation between \ax~and
$f_{\rm x}/f_{\rm r}$ for HBLs, significant at the 99.99 per cent level
(Spearman rank-order test), independently of the precise value of the dividing
$f_{\rm x}/f_{\rm r}$ ratio, and not present in LBLs. A linear fit to the HBL
data gives $\alpha_{\rm x} = -(0.41\pm0.07) \log (f_{\rm x}/f_{\rm r}) -
(2.65\pm0.66)$. In HBLs, in other words, the smaller the X-ray to radio flux
ratio, (i.e., the steeper the effective radio-X-ray spectral index), the
steeper is the X-ray spectral index, which points to a connection between the
radio and X-ray bands and/or to the effects of a break in the energy 
distribution (see Sect. 3.3). 

\subsection{\ax~versus $\alpha_{\rm ox}$}  

A similar and related connection for HBLs exists between the optical and X-ray
bands. Figure 4 plots the X-ray spectral slope versus $\alpha_{\rm ox}$, the 
effective optical-X-ray spectral index evaluated between the rest-frame 
frequencies of 5000 \AA~and 1 keV. Optical fluxes were derived from the V
magnitudes collected by Padovani \& Giommi (1995b), correcting for Galactic
absorption following Giommi et al. (1995a), while 1 keV fluxes have been
obtained from the {\it ROSAT} counts and the derived spectral indices and have
also been corrected for Galactic absorption. The $k$-correction has been 
derived assuming an optical index $\alpha_{\rm o} = 1.05$ for LBLs and 
$\alpha_{\rm o} = 0.65$ for HBLs (Falomo, Scarpa \& Bersanelli 1994). For
$k$-correction purposes objects with no redshift have been assigned the mean
redshift of the corresponding sub-sample, i.e. $z_{\rm mean, LBL} = 0.60$ and
$z_{\rm mean, HBL} = 0.25$. Note that optical fluxes are not simultaneous with
X-ray data (although for most of the 1 Jy BL Lacs in our sample they should
be representative of their ``typical'' optical state: see Padovani \& Giommi
1995b), which will certainly introduce a scatter: a variation $\Delta V$ in
magnitude, in fact, translates into a change $\Delta \alpha_{\rm ox} = 0.15
\Delta V$. 

\beginfigure{4}
\psfig{file=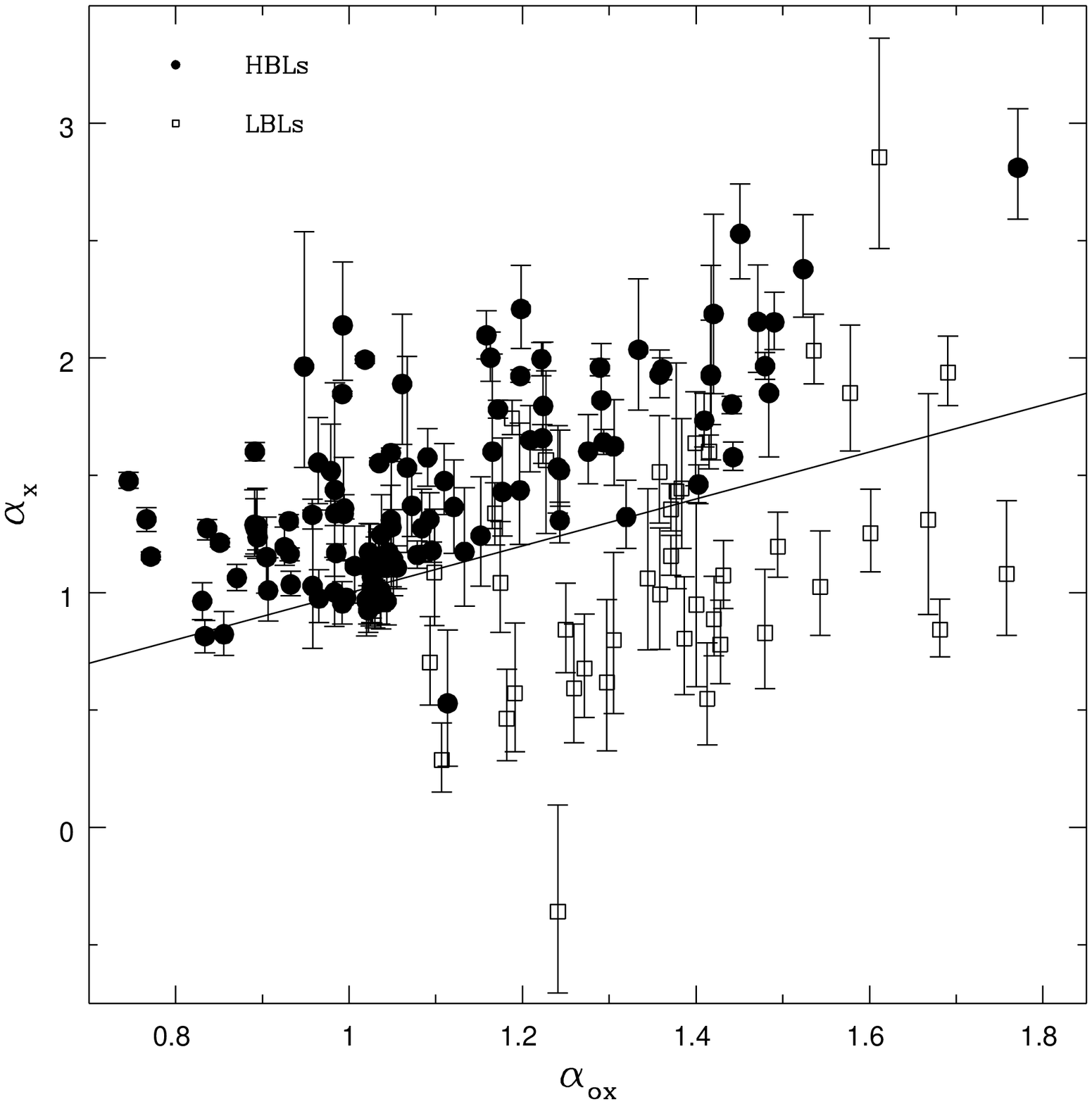,height=8.5truecm,width=8.5truecm}
\caption{{\bf Figure 4.} The X-ray spectral index versus $\alpha_{\rm ox}$ for
BL Lacs. Filled points indicate HBLs ($f_{\rm x}/f_{\rm r} \ge
10^{-11.5}$), while open squares indicate LBLs ($f_{\rm x}/f_{\rm r} <
10^{-11.5}$). The solid line represents the locus of points having
\ax~$=\alpha_{\rm ox}$. Error bars represent $1\sigma$ errors.} 
\endfigure

From the data presented in Fig. 4 we infer that: 1. for HBLs the X-ray
spectral index is correlated with $\alpha_{\rm ox}$ at the 99.99 per cent
level (Spearman rank-order test) with a best fit $\alpha_{\rm x} =
(1.38\pm0.20) \alpha_{\rm ox} - (0.01\pm0.22)$; 2. no correlation is present
for LBLs; 3. the two classes separate extremely well on this plane, that is
all HBLs apart from one have $\alpha_{\rm x} \ge \alpha_{\rm ox}$ (within the
errors), while the majority of LBLs have $\alpha_{\rm x} < \alpha_{\rm ox}$. 

This separation is better appreciated in Fig. 5, which plots $\alpha_{\rm x} -
\alpha_{\rm ox}$ versus $f_{\rm x}/f_{\rm r}$. This plane can be divided into 
four quadrants, out of which one is basically unpopulated and another one is 
only marginally populated. The remaining ones are the HBL--convex-spectrum
quadrant ($f_{\rm x}/f_{\rm r} \ge 10^{-11.5}$ and $\alpha_{\rm x} >
\alpha_{\rm ox}$) and the LBL--concave-spectrum quadrant ($f_{\rm x}/f_{\rm r}
< 10^{-11.5}$ and $\alpha_{\rm x} < \alpha_{\rm ox}$). The only HBL with 
possibly concave spectrum is MS0122.1+0903, which is however consistent with
having $\alpha_{\rm x} > \alpha_{\rm ox}$ at the $\sim 2\sigma$ level,
especially considering that errors on $\alpha_{\rm x} - \alpha_{\rm ox}$ do 
not include errors on $\alpha_{\rm ox}$. As regards the twelve LBL spectra with
possibly convex optical-X-ray continuum, only four of them have $\alpha_{\rm
x} > \alpha_{\rm 
ox}$ at the $\ga 2\sigma$ level. These correspond to four different objects:
one is 1WGA J1202.1+444, discussed above, while the other three have $f_{\rm
x}/f_{\rm r}$ values within a factor of two of the HBL/LBL dividing line,
which, as far as we know, could have well been crossed had there been a
decrease in radio flux by a factor of $2$ (compared to the values we used)
at the time of the {\it ROSAT} observation. 

\beginfigure{5}
\psfig{file=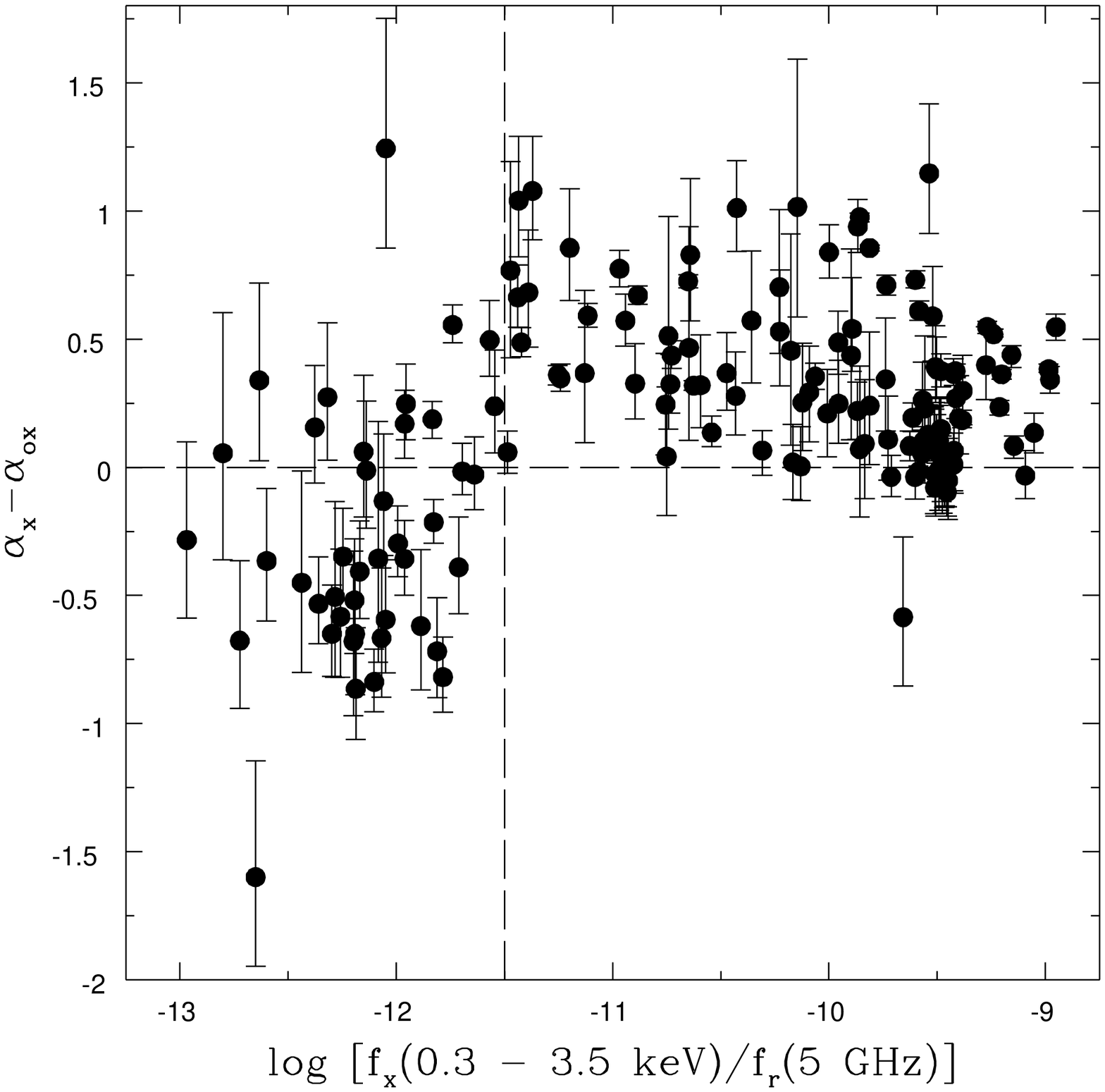,height=8.5truecm,width=8.5truecm}
\caption{{\bf Figure 5.} \ax~minus $\alpha_{\rm ox}$ versus $f_{\rm x}/f_{\rm
r}$ for BL Lacs. The horizontal dashed line represents the locus of points
with \ax~$=\alpha_{\rm ox}$, while the vertical dashed line represents the
dividing line between LBLs and HBLs at $f_{\rm x}/f_{\rm r} = 10^{-11.5}$.
Error bars represent $1\sigma$ errors and include {\it only} the uncertainties
on \ax.} 
\endfigure

The difference between HBLs and LBLs in $\langle \alpha_{\rm x} - \alpha_{\rm
ox}\rangle$ is striking, this parameter being equal to $0.40\pm0.04$ for HBLs
and $-0.33\pm0.09$ for LBLs ($0.17\pm0.06$ for the whole sample). This is
simply an amplification of the different X-ray and effective optical-X-ray
spectral indices. While, as we have 
seen, HBLs have \ax~steeper than LBLs, the opposite is true as regards
$\alpha_{\rm ox}$ (see Fig. 4), for which $\langle \alpha_{\rm ox,HBL} \rangle
= 1.11\pm0.03$ and $\langle \alpha_{\rm ox,LBL} \rangle = 1.39\pm0.03$. 

Sambruna (1994; see also Maraschi \ea 1995) has found a significant difference
between the $\alpha_{\rm ox} - \alpha_{\rm x}$ distributions of 1 Jy RBLs and
EMSS XBLs, with mean values, however, apparently less different than ours.
This difference can be explained by her smaller statistics and inclusion of
three 1 Jy HBLs with the RBL sample. 

Perlman \ea (1995b) have compared the {\it ROSAT} X-ray spectral indices of 22
EMSS BL Lacs (all HBLs) to the slope of the tangent at 1 keV to a parabola
fitted to the radio, optical, and X-ray fluxes. They find that the X-ray
spectra have similar or slightly steeper slopes than the tangent slope in most
cases, a somewhat different result from ours. This might be due to our larger
sample and also to their use of the tangent slope instead of the effective
optical-X-ray index. Brunner \ea (1994) have studied the {\it ROSAT} spectra
of five 1 Jy BL Lacs and derived a mean value for $\alpha_{\rm x} -
\alpha_{\rm ox}$ consistent with zero, unlike our result for LBLs. Despite
their small statistics, exclusion of the HBL object in their sample (S5
$0716+714$), which has the largest value of $\alpha_{\rm x} - \alpha_{\rm
ox}$, makes their mean value consistent with ours. 

\subsection{\ax~versus $\nu_{\rm break}$}  

The anti-correlation between \ax~and $f_{\rm x}/f_{\rm r}$ and the correlation
between \ax~and $\alpha_{\rm ox}$ for HBLs are both related to what we think
is the most interesting correlation, that between \ax~and $\nu_{\rm break}$.

In theory, one should infer the break frequency $\nu_{\rm break}$ from the
observed energy distribution of the objects. In practice, however, the
available spectral coverage is far from complete and the measurements are
generally non-simultaneous. In particular, far-ultraviolet and far-infrared
data are missing for the large majority of the sources, and these are the
bands where the peak of the emission is most likely located for HBLs and LBLs
respectively (Giommi \ea 1995a). The break frequency has then been derived as
follows: Padovani \& Giommi (1995a), using a triple power-law approximation to
the spectra, have shown that the distribution of BL Lacs on the $\alpha_{\rm
ro}, \alpha_{\rm ox}$ plane (where $\alpha_{\rm ro}$ is the effective
radio-optical spectral index evaluated between the rest-frame frequencies of 5
GHz and 5000 \AA) can be easily explained by an energy break migrating from
high (X-ray) to low (far-infrared) energies (see their Fig. 12). Here we use
their results to infer $\nu_{\rm break}$ from the $\alpha_{\rm ro}$ and
$\alpha_{\rm ox}$ values. The former were derived from the radio and optical
data collected by Padovani \& Giommi (1995a), using optical indices and mean
redshifts (when necessary) as described above. As regards the radio spectral
index, necessary for the $k$-correction of $\alpha_{\rm ro}$, for the BL Lacs
lacking this parameter it was assumed equal to the mean value for the two
sub-classes, that is $\alpha_{\rm r} = 0.2$ for HBLs and $\alpha_{\rm r} =
-0.2$ for LBLs. 

Operationally, we projected the points on the $\alpha_{\rm ro}, \alpha_{\rm
ox}$ plane on the predicted relationship between the two parameters derived
by Padovani \& Giommi (1995a), namely on that of HBLs for $f_{\rm x}/f_{\rm r}
\ge 10^{-11.5}$ and on that of LBLs for $f_{\rm x}/f_{\rm r} < 10^{-11.5}$ 
(see Fig. 12 of Padovani \& Giommi 1995a). We then read off the corresponding 
values of the break frequency. Although this method gives only an approximate
estimate of $\nu_{\rm break}$, the values derived this way agree well with
those inferred from an inspection of the plots, for example, in Giommi et al.
(1995a) for the (few) objects with enough data to permit a direct
determination of the break. Moreover, our values are typically consistent
within $\sim 0.2$ dex with the peak frequencies obtained by Comastri \ea
(1995) from polynomial fits to the broad band spectra of twelve 1 Jy BL Lacs. 

\beginfigure{6}
\psfig{file=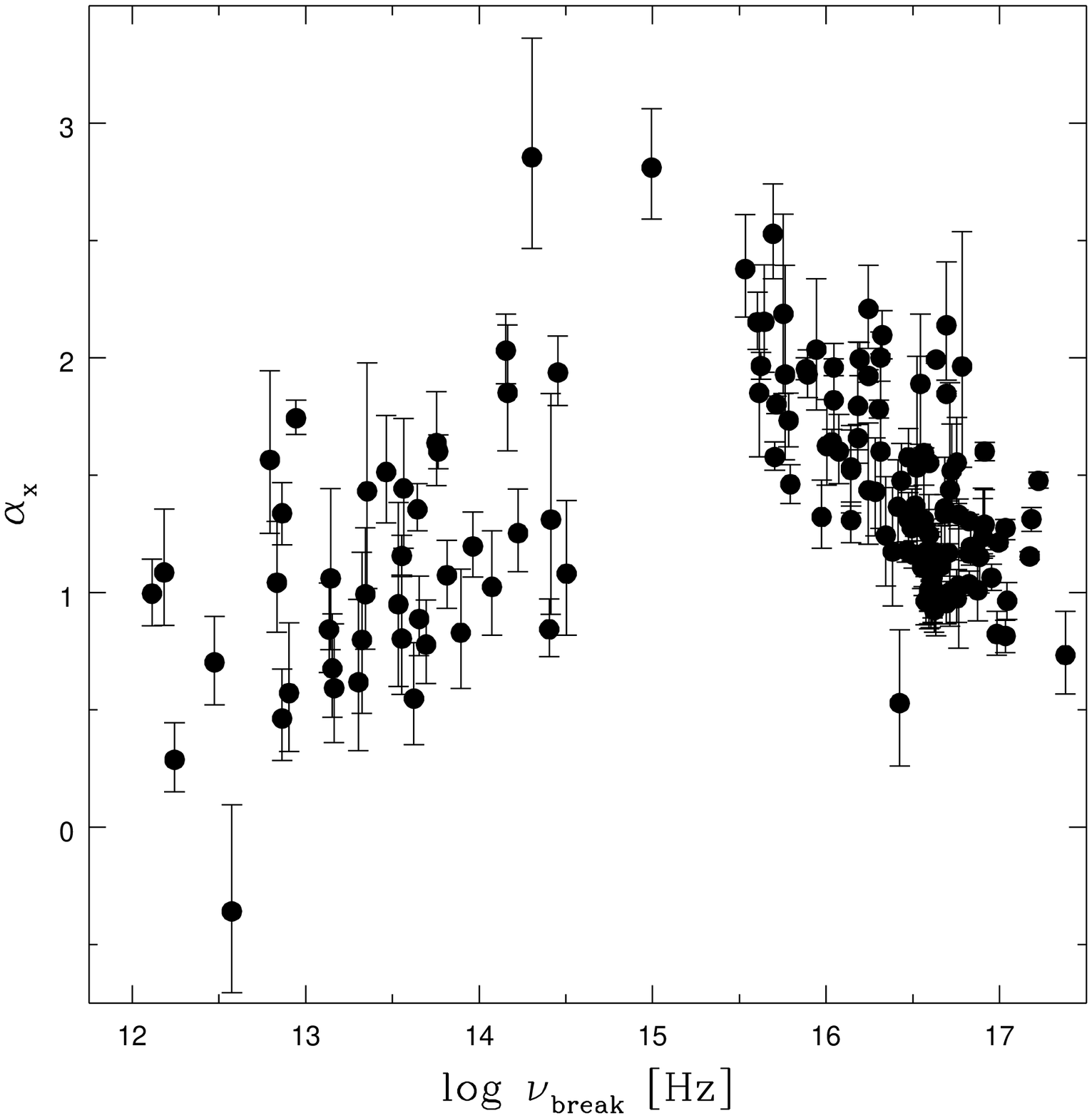,height=8.5truecm,width=8.5truecm}
\caption{{\bf Figure 6.} The X-ray spectral index versus the logarithm of the 
break frequency for BL Lacs. Objects with $\nu_{\rm break} \ga 10^{15}$ Hz have
$f_{\rm x}/f_{\rm r} > 10^{-11.5}$ and are, by definition, HBLs, while the 
remaining objects are LBLs. Error bars represent $1\sigma$ errors. See text
for the derivation of $\nu_{\rm break}$.} 
\endfigure

Figure 6 clearly shows the strong anti-correlation (significant at the 99.99
per cent level: Spearman rank-order test) between X-ray spectral index and
break frequency for HBLs. A linear fit gives $\alpha_{\rm x} =
-(0.63\pm0.09)\log \nu_{\rm break} + (11.87 \pm 1.50)$: i.e., the highest
$\nu_{\rm break}$, the flattest \ax. As regards LBLs, here the situation is
not clear: a fit to the ``compressed'' data as done for all the correlations
studied in this paper (i.e., using only one spectrum per object, selected as
described in Sect. 2) gives no correlation between \ax~and $\nu_{\rm break}$
for the 27 LBLs ($P{\rm corr} \simeq 61$ per cent). However, Fig. 6 shows an
apparent 
positive correlation between the two parameters ($\alpha_{\rm x} \propto 0.41
\log \nu_{\rm break}$) when all (42) spectra are included, significant at the
99.6 per cent level (Spearman rank-order test; this reduces to 98.2 per cent,
with $\alpha_{\rm x} \propto 0.26 \log \nu_{\rm break}$, if one excludes the
two outliers with the largest and smallest spectral index). This might suggest
that LBLs display a (weak) positive correlation, possibly diluted by the 
relatively small statistics and larger error bars (the LBL sample is less than
half the size the HBL one and on average the LBLs studied in this work are
weaker X-ray sources than HBLs). 

\subsection{Interpretation}

The negative correlation between \ax~and $\nu_{\rm break}$ observed for HBLs
is expected if the emission of these objects is due to the synchrotron process
all the way up to the X-ray band, as shown in Fig. 7: when the break is near
the X-ray band ($\sim 10^{17}$ Hz: dotted line) the X-ray spectrum will be
relatively flat, but as soon as the peak shifts to lower energies ($\sim
10^{16}$ Hz: dashed line) the steep tail of the synchrotron emission moves in.
When $\nu_{\rm break}$ is at much lower energies (e.g., $\sim 10^{14}$ Hz:
solid line) so that the synchrotron component is no more dominant, Compton
emission starts to be important: at this point one has crossed the HBL/LBL
dividing line. For even smaller break frequencies, two effects compete: first,
the smaller $\nu_{\rm break}$, the smaller is the product $B \gamma_{\rm
max,e}^2$, where $B$ is the magnetic field and $\gamma_{\rm max,e}$ is the
maximum Lorentz factor of the electrons (e.g., Rybicki \& Lightman 1979); this
should not affect the lower energies, thereby leaving unchanged the shape of
the Compton component; second, a lower $\nu_{\rm break}$ means a weaker steep
X-ray synchrotron component, which should result in a flatter spectrum in
objects with still a non-negligible synchrotron emission. The sum of the two
effects is not easy to quantify but the fact that in LBLs the X-ray spectral
index is not strongly dependent on $\nu_{\rm break}$ suggests that the Compton
component might dominate completely the X-ray band in at least some objects. 

This ties in nicely with the study of the $\alpha_{\rm ox}$ index, which shows
that the X-ray band connects smoothly to the optical one in HBLs, with a
steepening of the broad-band spectrum, supporting a common emission mechanism
for the two bands, while in the great majority of LBLs \ax~is flatter than
$\alpha_{\rm ox}$, indicative of a different component entering the X-ray 
range. Also, the anti-correlation between \ax~and $f_{\rm x}/f_{\rm r}$ in HBLs
is simply related to that between \ax~and $\nu_{\rm break}$ as objects with 
larger $\nu_{\rm break}$ have also larger values of $f_{\rm x}/f_{\rm r}$.  

In objects with very low break frequencies, for which the X-ray emission
should be pure inverse Compton, the X-ray spectral indices should be similar
to those in the radio--far-infrared region, obviously up to $\nu \la \nu_{\rm
break}$. Unfortunately, very few BL Lacs have been detected by {\it IRAS} so
the statistics is quite poor: we find that the 5 GHz -- 60$\mu$ spectral index
for about ten 1 Jy BL Lacs is $\alpha_{\rm ri} \sim 0.3$ (see also Giommi \ea
1995a), which roughly corresponds to the lower bound of \ax~for LBLs. Clearly
a detailed study of the far infrared spectra of LBLs would be extremely
valuable in this respect: this project will be carried out with {\it ISO}. 

\beginfigure{7}
\psfig{file=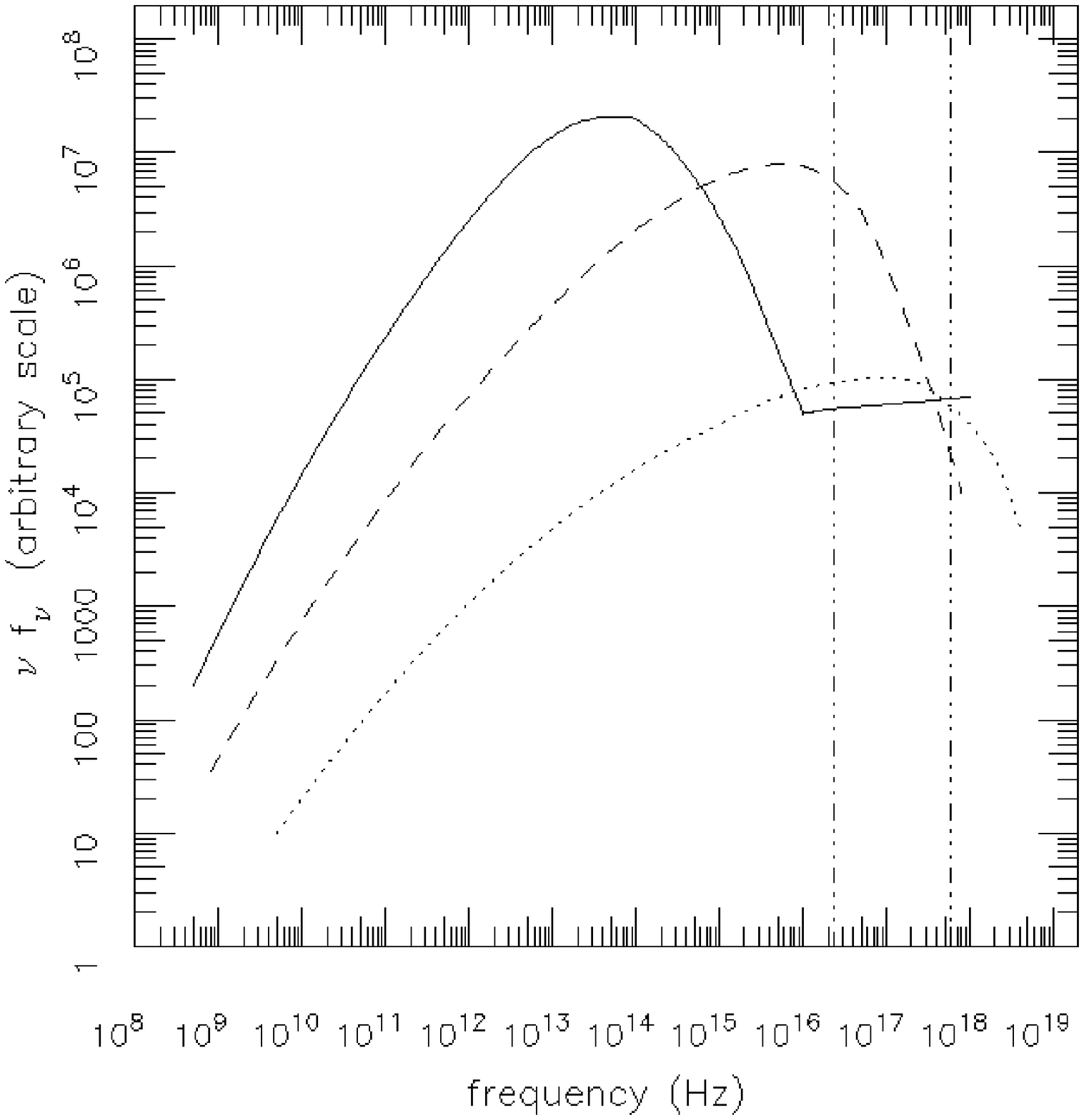,height=8.5truecm,width=8.5truecm}
\caption{{\bf Figure 7.} Examples of multifrequency spectra of BL Lacs with
different peak (or break) frequencies: $\sim 10^{17}$ Hz ($\sim 0.4$ keV:
dotted line), $\sim 10^{16}$ Hz (dashed line), $\sim 10^{14}$ Hz (solid line).
The two parallel (dot-dashed) lines represent the {\it ROSAT} band. Note how
the X-ray spectrum steepens with decreasing break frequency until Compton
emission becomes the dominant process. The normalizations of the different
curves have been chosen to best illustrate the spectral differences in the 
X-ray band and do not reflect intrinsic properties.} 
\endfigure

We predict that the observed anti-correlation between \ax~and $\nu_{\rm
break}$ observed for HBLs cannot continue indefinitely. Since the synchrotron
spectrum steepens with frequency (e.g., Ghisellini \& Ma\-ra\-schi 1989), if
$\nu_{\rm break}$ moves to larger and larger values, the X-ray spectral index
at minimum will be as flat as the effective radio-optical spectral index
$\alpha_{\rm ro}$. In that case, then, one will have $\alpha_{\rm x} \sim
\alpha_{\rm ox} \sim \alpha_{\rm ro} \sim 0.3 - 0.5$, the typical values for
HBLs. Extrapolating the observed anti-correlation, one should find
$\alpha_{\rm x} \sim 0.4 \sim$ constant for $\nu_{\rm break} \ga 10$ keV. As
this requires very high electron energies, it is not surprising that no BL Lac
in our sample has such a high $\nu_{\rm break}$. If such objects exist they
are probably quite rare but should be detected by future hard-X-ray missions. 

We note that the paucity of objects with break frequencies in the
optical/ultraviolet band ($\sim 10^{14.5} - 10^{15.5}$ Hz) is simply a
selection effect, as these objects have intermediate values of $f_{\rm
x}/f_{\rm r}$ which make them difficult to detect both in radio and X-ray
surveys. Optical surveys, which in principle do not discriminate against this
sort of objects, are at present very inefficient (e.g., Green, Schmidt \&
Liebert 1986; Jannuzi, Green \& French 1993) and have so far led to the
selection of only about ten BL Lacs (Padovani \& Giommi 1995b). 

\section{Discussion and conclusions}

The data presented in this paper allow us to draw a consistent picture of HBLs
and LBLs. Namely, the correlations of the X-ray spectral indices with 
$\alpha_{\rm ox}$, $f_{\rm x}/f_{\rm r}$, and break frequency all point to a
common, likely synchrotron, origin for the radio/optical/X-ray emission in
HBLs, with an overall convex spectrum. On the other hand, as hinted by the
lower cutoff energies, in LBLs a second, harder, most likely inverse Compton,
component appears in the X-ray band. The {\it ROSAT} spectral indices of the
two classes are different, with HBLs displaying on average steeper spectra
which show a strong anti-correlation with break frequency, while LBL spectra
might show a {\it positive} dependence on this parameter. These are yet other
differences between the two BL Lac classes, which add to the previously known
ones, and might be helpful in selecting BL Lac candidates from the WGA 
catalogue, for example. 


Based on the additional information presented here, we can better pinpoint the
borderline between HBLs and LBLs, and therefore propose a refined dividing
line between the two classes 
at $f_{\rm x}/f_{\rm r} \sim 10^{-11.5}$, which corresponds to $\alpha_{\rm
rx} \sim 0.8$ between 1 keV and 5 GHz and a break frequency $\sim 10^{15}$
Hz (see Fig. 6). Our results agree with the hypothesis we put forward that the
cutoff energy is the fundamental parameter giving rise to two different
classes of BL Lacs (Giommi \& Padovani 1994; Padovani \& Giommi 1995). Oddly
enough, our results are also consistent with the other explanation for the
existence of two BL Lac classes based on orientation. Ghisellini \& Maraschi
(1989), in fact, have shown that in objects at larger angles with respect to
the line of sight (i.e., XBLs/HBLs in their view) synchrotron emission should
be the main radiation mechanism in the X-ray band, while inverse Compton
should be more important for more aligned sources (i.e., RBLs/LBLs), in
agreement with our findings. The main problem of that model, however, apart
from other ones discussed by Padovani \& Giommi (1995a), is that a change in
viewing angle does not seem to be able to explain the large difference in
cutoff energies between the two classes, as exemplified by Fig. 4 of
Ghisellini \& Maraschi (1989) (Sambruna 1994; Sambruna \ea 1995, in
preparation). 

Our results explain the peculiar spectral variability of XBLs (i.e., HBLs) and
RBLs (i.e., LBLs). Various studies have shown that in XBLs higher flux states
are associated with flatter spectra (e.g., Giommi \ea 1990; Sambruna \ea
1994). Little is known about spectral variability in RBLs but Urry \ea (1996)
have shown that the handful of 1 Jy BL Lacs with suitable data seem to show
the opposite trend, that is steeper spectra in brighter states (the exception,
PKS $2005-489$, is an HBL). If X-ray variability in BL Lacs is accompanied by
a change in break frequency (e.g., George, Warwick \& Bromage 1988: MKN 421;
Sambruna 1994: S5 
0716+714, PKS 2005$-$489), this behaviour is easily explained: when an object
brightens and the break moves to higher energies, in HBLs the spectrum
hardens, as shown in Fig. 8 for a single source and in Figs. 2 and 6 for the
HBL population. The opposite effect for RBLs can be explained in those cases
where synchrotron emission in the X-ray band is not negligible in the high 
state: when the break moves to higher energies, the steep synchrotron
component moves in the X-ray band therefore steepening the spectrum (Fig. 8).
This would be in agreement with the (weak) correlation between \ax~and
$\nu_{\rm break}$ for LBLs, discussed in Sect. 3.3 (see Fig. 6). Figure 8 also
shows that a spectral variability similar to that observed in the X-ray band
for HBLs, that is flatter spectra at higher states, should be mirrored by LBLs
in the infrared band, as in fact observed (e.g., Gear \ea 1986). 


\beginfigure{8}
\psfig{file=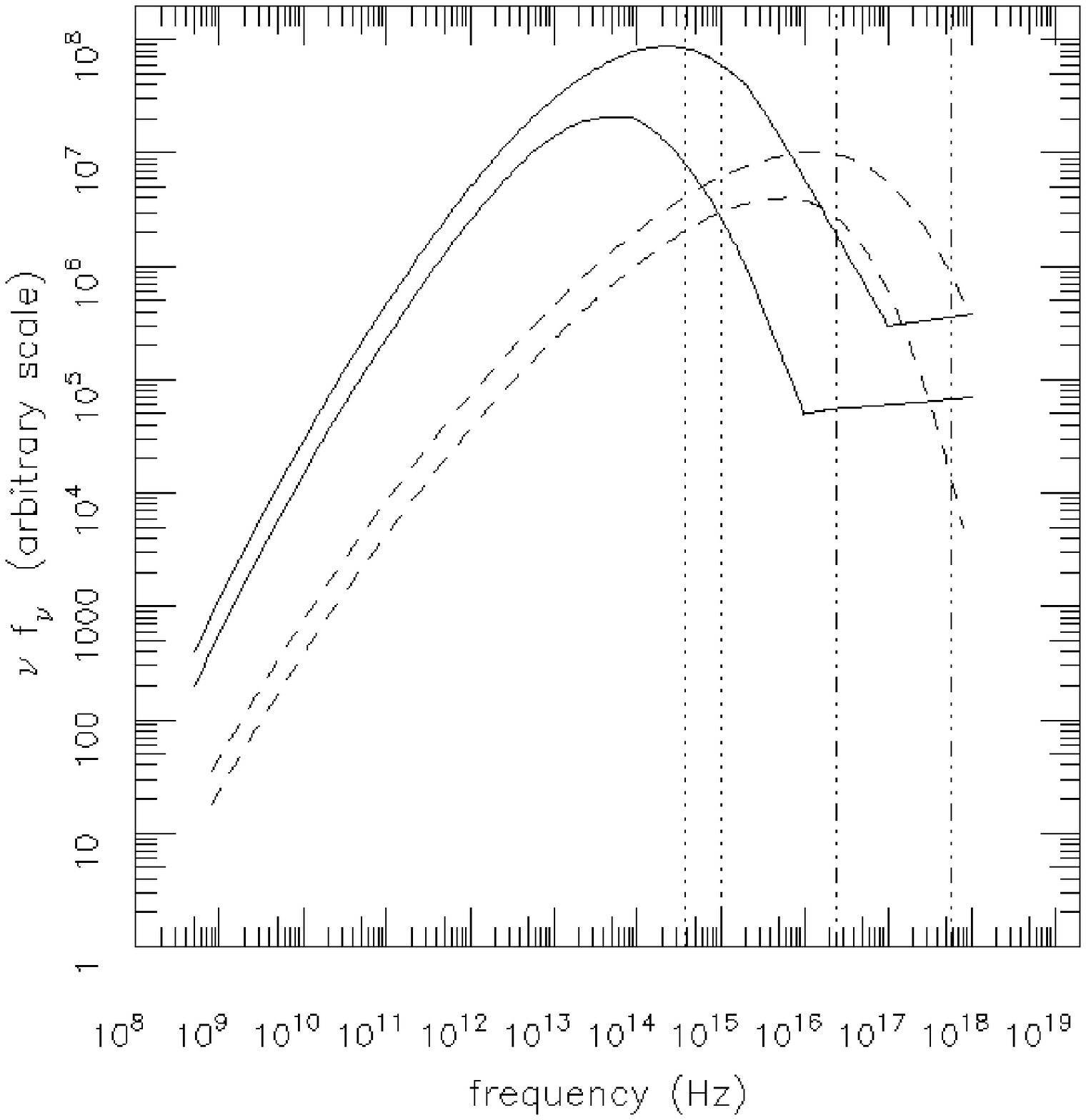,height=8.5truecm,width=8.5truecm}
\caption{{\bf Figure 8.} The effect of a change in break frequency on the flux
and spectral variability of BL Lacs: in HBLs (dashed lines), when the object
brightens and the break moves to higher energies, the X-ray spectrum hardens.
In LBLs (solid lines), the opposite effect might apply (i.e., spectral
steepening at higher fluxes) when synchrotron emission in the X-ray band is
not negligible in the high state. Incidentally, note also the much larger flux
variability, at fixed radio variability, expected for LBLs in the optical band
(parallel dotted lines), as compared to that of HBLs, as effectively observed
(e.g., Jannuzi \ea 1994). As in Fig. 7, the normalizations of the different
curves have been chosen only for illustrative purposes and do not reflect
intrinsic properties.} 
\endfigure

Our results seem in contradiction with those of Urry \ea (1996), who found
that the \ax~distribution for 1 Jy RBLs was consistent with that of the EMSS
XBLs. As discussed in the Introduction, however, most previous work on the
study of spectral indices of BL Lacs was limited by smaller statistics and the
use of the RBL/XBL classification, which can be misleading. In particular, the
result of Urry \ea (1996) can be explained as follows: 1. three of the 1 Jy
RBLs are HBLs (S5 $0716+714$, MKN 501, PKS $2005-489$), whose inclusion in the
RBL sample certainly shifted the \ax~distribution to steeper values (the mean
value of \ax~for these three objects is significantly different, i.e.,
steeper, than that of the rest of the sample: see also Comastri \ea 1995 and 
Urry \ea 1996); 2.
the EMSS BL Lacs (all HBLs) have values of $f_{\rm x}/f_{\rm r} \ga
10^{-10.5}$ (Padovani \& Giommi 1995a), partly because, being a serendipitous
survey, the EMSS excludes by definition well-know BL Lacs with considerable
radio emission which have lower $f_{\rm x}/f_{\rm r}$ values but more
importantly because objects with relatively low values of $f_{\rm x}/f_{\rm
r}$ have a small probability of being detected and large-area X-ray surveys
are needed. (This interpretation is supported by the fact that the Slew survey
HBLs reach lower $f_{\rm x}/f_{\rm r}$ values: see Perlman et al. 1995a).
Given the anti-correlation between \ax~and $f_{\rm x}/f_{\rm r}$ (Fig. 2),
this implies that EMSS HBLs are bound to have \ax~flatter than average,
thereby reducing any difference between the two classes. In fact, if we
compare our LBL sample with the EMSS HBLs, we also find no significant
difference between the spectral indices of the two samples. 

One could argue that our sample is not complete in a statistical sense.
However, it is the largest sample for which homogeneous X-ray data are
available and includes 50 per cent of known BL Lacs. We needed a large enough
sample to maximize the coverage of the parameter space to be able to test for
any differences in the X-ray spectra of the two BL Lac classes. Previous
results, even those based on complete (and therefore not very large) samples,
were in fact not clear-cut. We are not aware of selection effects which could
possibly alter our results significantly. Furthermore, considering only the
complete sub-sample of 22 EMSS BL Lacs (all HBLs: Perlman \ea 1995b) we still
reproduce one of our most important results, i.e. we find an anti-correlation
between \ax~and $\nu_{\rm break}$, significant at the 99.3 per cent level
(Spearman rank-order test), consistent with the one we obtained for the whole
HBL sample. 

Comastri \ea (1995) have found a {\it positive} correlation between \ax~and
$\nu_{\rm break}$ in their sample of twelve 1 Jy BL Lacs (which includes nine
LBLs and three HBLs). This is consistent with our results. If we consider all
objects irrespective of the HBL/LBL classification, in fact, we also detect a
positive correlation between \ax~and $\nu_{\rm break}$, because on average
HBLs have steeper spectra. Our much better statistics, however, allows us to
examine this relation in more detail and to see that the dependence of \ax~on
break frequency is different for the two BL Lac classes, in particular HBLs
exhibit a {\it negative} correlation between the two parameters. 

The main conclusions of this paper, which studies the {\it ROSAT} X-ray
spectra of about half the known BL Lacs, can be thus summarized: 
\medskip
BL Lacs are characterized by energy power-law spectral indices ranging between
$\sim 0$ and $3$, with an average value $\sim 1.4$. The X-ray spectral 
properties of BL Lacs with high ($\ga 10^{15}$ Hz) frequency cutoffs (HBLs)
are significantly different from those with lower frequency cutoffs (LBLs).
HBLs have on average steeper X-ray spectral indices (\ax~$\sim 1.5$) which
correlates with the effective radio-X-ray and optical-X-ray spectral indices.
The overall spectrum is convex from radio to X-ray energies, and gets steeper
with frequency. A strong anti-correlation is also present between X-ray
spectral index and break frequency, with sources having cutoffs at larger
energies displaying harder X-ray spectra. In LBLs, on the other hand, which
have \ax~$\sim 1.1$, the X-ray spectrum shows no correlation with the
effective spectral indices and the optical-X-ray continuum has a concave
shape. These differences might also explain the peculiar X-ray spectral
variability of HBLs and LBLs. 

These findings strongly support the hypothesis that the main radiation
mechanism in HBLs is synchrotron emission from the radio to the X-ray band,
while in LBLs an inverse Compton component dominates the X-ray emission, and
are consistent with the view that the apparent dichotomy between the two BL
Lac classes is mainly due to different cutoff energies. They would also be
consistent with the predictions of an inhomogeneous SSC model in which the
main difference is the orientation of the beaming axis, but this model appear
not to be able to reproduce the observed large differences in cutoff energies.

\section*{Acknowledgments}

We thank Fabrizio Fiore, Gabriele Ghisellini, and Rita Sambruna for useful
discussions, and Eric Perlman and Meg Urry for providing us with results in
advance of publication. This research has made use of the BROWSE program
developed by the ESA/EXOSAT Observatory and by NASA/HEASARC. We acknowledge
partial support from ESA/ESRIN, Frascati, where part of this work was done,
while P.G. was the ESIS senior scientist and P.P. was a visiting scientist. 


\bigskip
\leftline{\bf NOTE ADDED IN PROOF}
\medskip
After this paper was accepted for publication, it came to our knowledge that
Lamer, Brunner \& Staubert (1996), A\&A, in press, have independently analysed
most of the data presented in this paper, reaching similar conclusions. 

\end